\documentclass{ws-ijmpcs}

\begin{document}

\markboth{A. Ventura}
{Searches for supersymmetry}

%
\catchline{}{}{}{}{}
%

\title{Searches for supersymmetry}

\author{Andrea Ventura}

\address{Dipartimento di Matematica e Fisica ``E. De Giorgi'' dell'Universit\`a del Salento \\ 
\& Sezione di Lecce dell'Istituto Nazionale di Fisica Nucleare\\
Via per Arnesano, Lecce, I-73100, Italy.\\
E-mail: andrea.ventura@le.infn.it\\
\vspace{5mm}
on behalf the ATLAS and CMS Collaborations}

\maketitle

\begin{history}
\end{history}

\begin{abstract}
New and recent results on Supersymmetry searches are shown for the ATLAS and the CMS experiments. Analyses with about 36 fb$^{-1}$ are considered for searches concerning light squarks and gluinos, direct pair production of 3$^{rd}$ generation squarks, electroweak production of charginos, neutralinos, sleptons, R-parity violating scenarios and long-lived particles.
\keywords{SUSY, ATLAS, CMS.}
\end{abstract}

\section{Introduction}	
Supersymmetry (or SUSY) \cite{susy1,susy2,susy3,susy4} is one of the most studied and favoured extensions of the Standard Model (SM), which postulates that for every particle in the SM there exists a supersymmetric partner (or {\it sparticle}), differing by a half unit of spin from their SM counterparts, with the squarks ($\tilde{q}$) being the scalar partners to the SM quarks and the gluinos ($\tilde{g}$) the fermionic partners to the SM gluons.

If sparticles exist at the TeV scale, they can be accessed at the Large Hadron Collider (LHC) presently running at CERN (Geneva), whose center-of-mass energy has been increased from 7-8 TeV (with about 25 fb$^{-1}$ collected during {\it Run 1} in 2011-12) to 13 TeV (with 36 fb$^{-1}$ collected during {\it Run 2} in 2015-16 and additional statistics to be recorded and analyzed in 2017).

SUSY presents many advantages since it can solve a number of still open questions in the SM, by providing a solution to the hierarchy problem, by suggesting the lightest supersymmetric particle (LSP) as a possible candidate for Dark Matter in R-Parity \footnote{R-parity is defined for each (s)particle as $R = (-1)^{3(B-L)+2S}$, where $B$ is the barionic number, $L$ is the leptonic number and $S$ is the spin. $R$ is $+1$ for SM particles and $-1$ for their SUSY partners.} conserving (RPC) models and by allowing gauge coupling unification  at high energies, which is not possible in the SM at any scale.

\begin{figure}[pb]
\hspace{-13mm}
\centerline{\includegraphics[width=11cm]{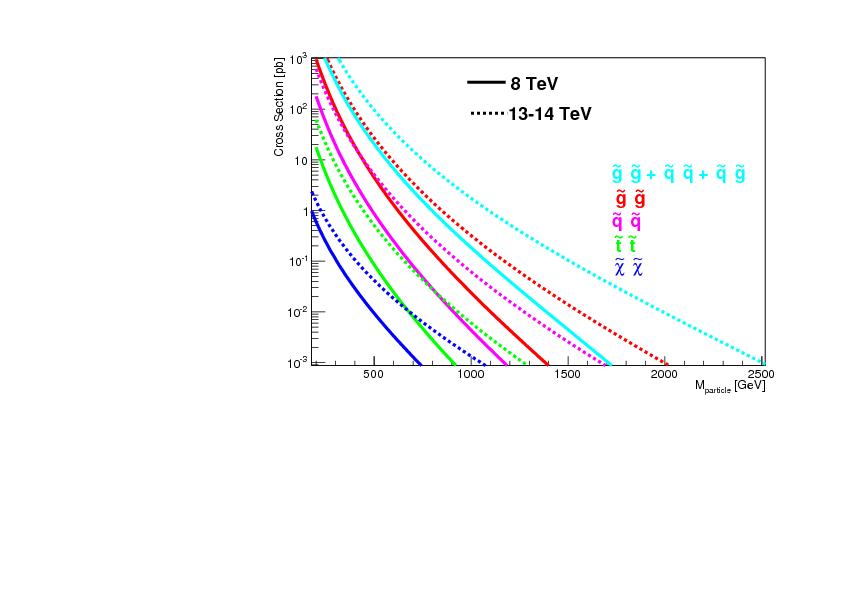}}
\vspace*{8pt}
\vspace{-26mm}
\caption{Cross sections for SUSY particle production at $\sqrt{s} = 8$ TeV and $13-14$ TeV.\label{crossections}}
\end{figure}

This work concentrates on an overview of results obtained in SUSY searches by the ATLAS \cite{atlasexp} and the CMS \cite{cmsexp} experiments running at the LHC.
Many models predicting sparticles observable at the LHC have been developed, ranging from easy-to-observe models with clear signatures and large cross sections to models with very low signal cross sections that are difficult to distinguish from the relevant SM backgrounds. The cross sections of the expected SUSY production mechanisms are shown in Fig. \ref{crossections} as functions of the produced sparticles. The represented colored particle cross sections are from {\it nll-fast} \cite{Beenakker2011fu} and evaluated at $\sqrt{s} = 8$ TeV and $13$ TeV, the electroweak pure higgsino cross sections are from {\it prospino} \cite{Beenakker1996ed} and evaluated at $\sqrt{s} = 8$ TeV and $14$ TeV.

Each of the analyses presented here uses selections on various observables in order to enhance signals with respect to SM background. Signal regions (SRs) are defined using Monte Carlo (MC) simulation of the signal and of the SM processes, suitable control regions (CRs) are used to quantify each background source, and validation regions (VRs) are defined and exploited to test the descriptions provided by such estimates.

\section{Inclusive squark and gluino searches}
\begin{figure}[pb]
\centerline{\includegraphics[width=7cm]{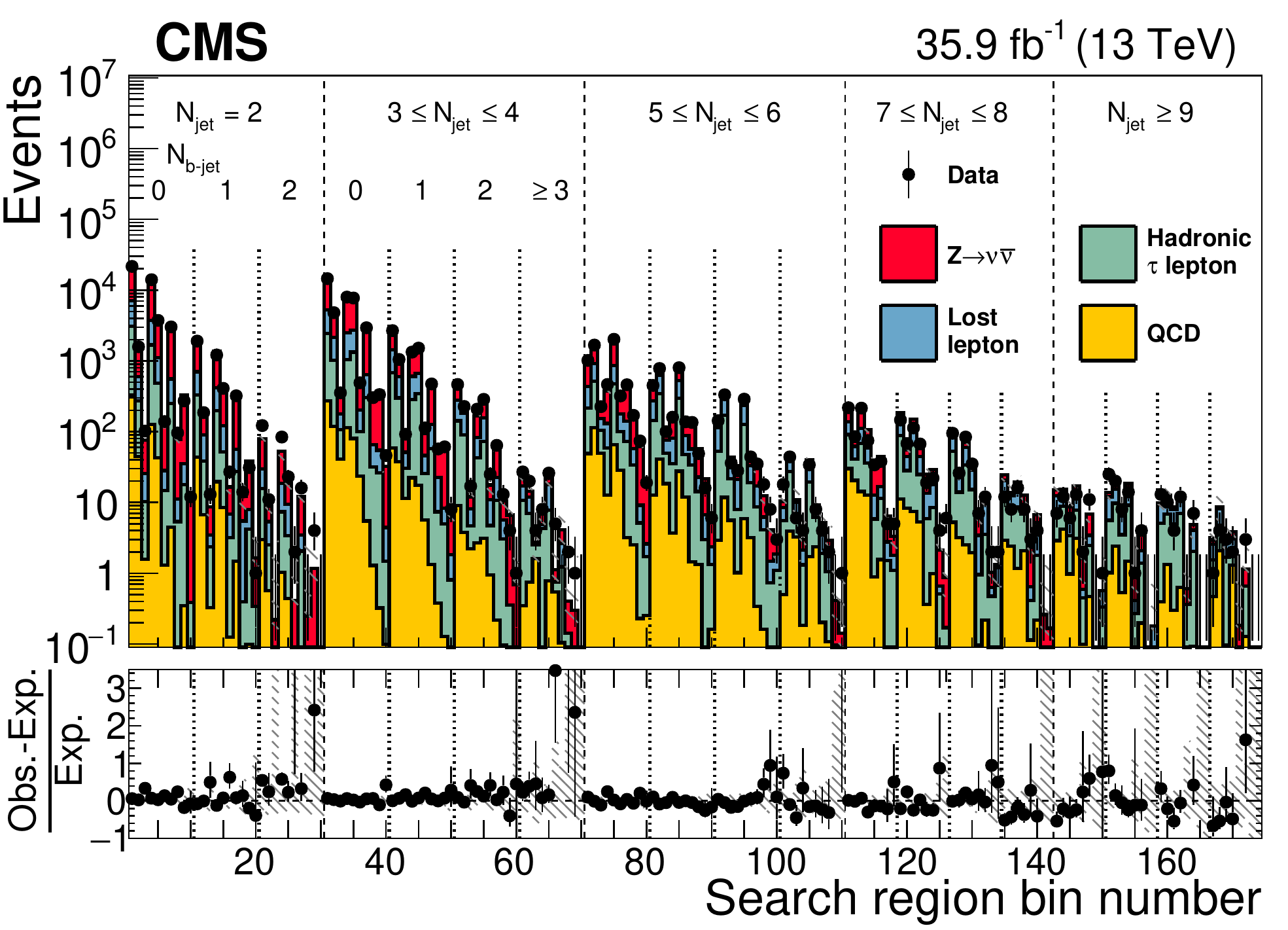}}
\vspace*{8pt}
\caption{Observed numbers of events and prefit SM background predictions in 174 different signal regions of the CMS searches for SUSY in multijet events with missing transverse momentum. \label{CMS0lep}}
\end{figure}

ATLAS and CMS have optimized a large number of analyses to search for simplified scenarios in which $\tilde{g}\tilde{g}$, $\tilde{q}\tilde{q}$ and $\tilde{q}\tilde{g}$ pairs are produced.
As an example, CMS searches for fully hadronic final states \cite{CMS_0lep} rely on a huge number of SRs, defined according to requirements on the number of ($b$-)jets in the event and on other variables, such as the missing transverse energy ($E_T^{miss}$) and the sum of jets transverse momentum $p_T$ ($H_T$). As shown in Fig.~\ref{CMS0lep}, the observed and expected SM event yields compare very well for all the considered SRs. Also the SUSY searches performed by ATLAS on decays with no leptons in the final state \cite{ATLAS_0lep} have shown no significant excess of events with respect to the SM expectation, as the largest significance measured in a SR corresponds to only 2.14 standard deviations.

\begin{figure}[pb]
\centerline{\includegraphics[width=8cm]{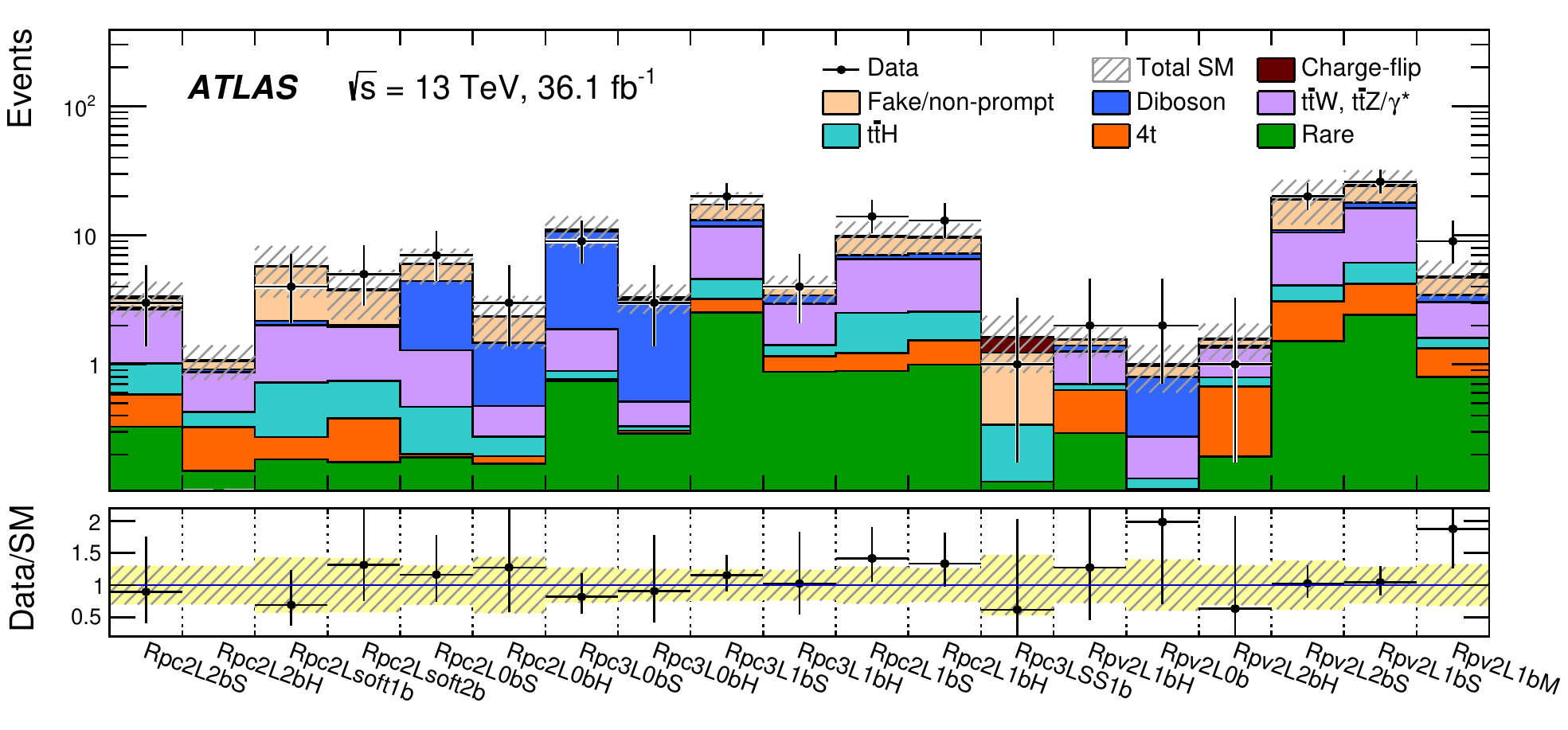}}
\vspace*{8pt}
\caption{Comparison of the observed and expected event yields in 19 different signal regions of the ATLAS searches for SUSY in events with two same-sign or three leptons and jets in the final state. \label{ATLAS2lSS}}
\end{figure}

Searches with 1, 2 or more isolated leptons in the final state present the advantage of removing the SM multijet background from the SRs. As an example, in Fig.~\ref{ATLAS2lSS} the event yield is shown for ATLAS searches \cite{ATLAS_2lSS} with two same-sign (or three) leptons in the final state. Owing to their special topology, the SRs are almost SM background free and the main limitation in the analysis is given by the reduced statistics. In the $\tilde{g}\tilde{g}$ simplified RPC models considered, the hypothesis of gluinos with masses up to 1.87 TeV is excluded in scenarios with a light $\tilde{\chi}^0_1$.

\begin{figure}[!tbp]
  \centering
  \begin{minipage}[b]{0.45\textwidth}
    \includegraphics[width=\textwidth]{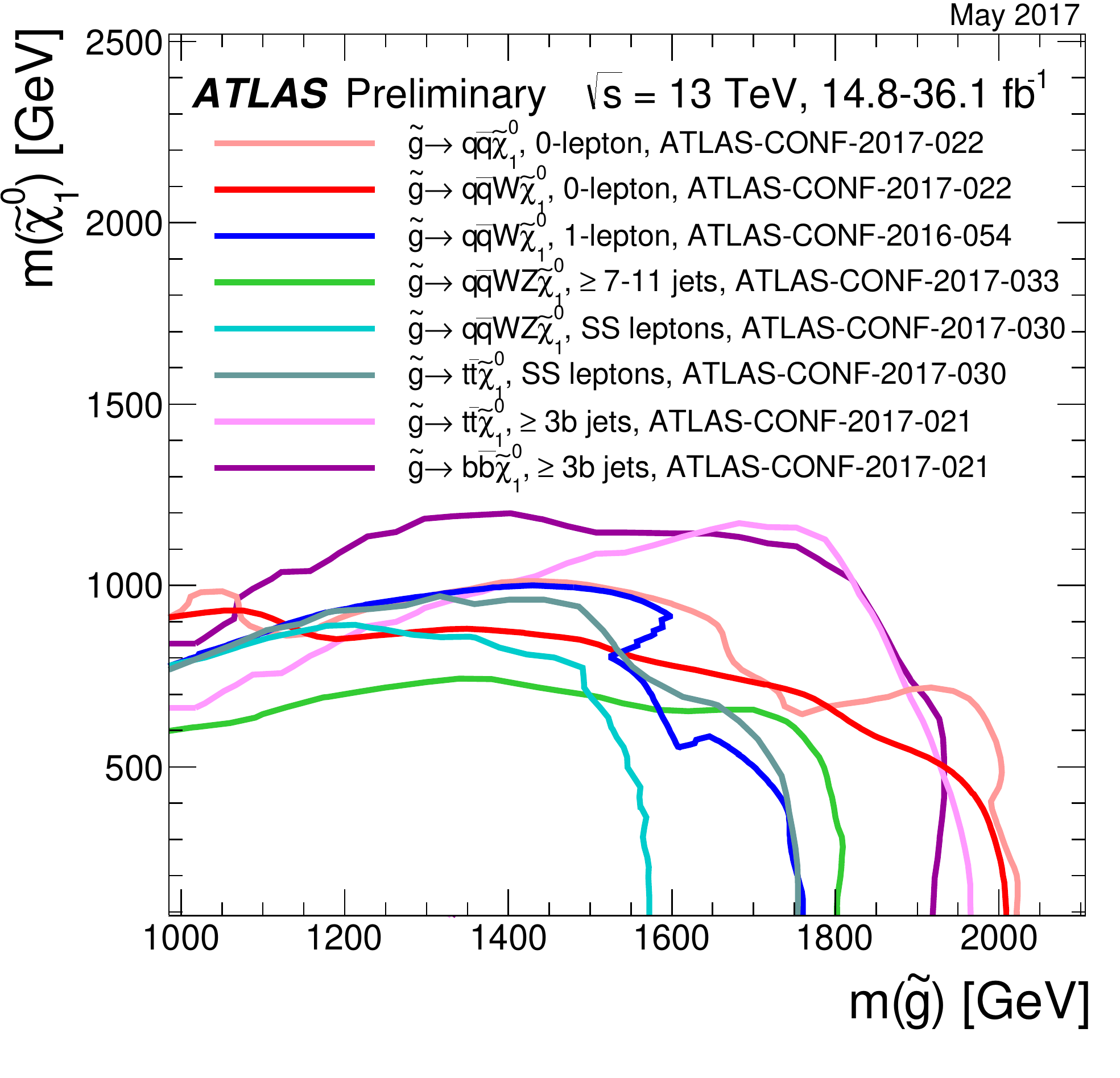}
	\caption{Summary of limits on gluino pair production with gluino decaying in different channels, based on 14.8 to 36.1 fb$^{-1}$ of pp collision data taken by ATLAS at $\sqrt{s}$ = 13 TeV. \label{ATLASstrong}}
  \end{minipage}
  \vspace{-1mm}
  \hfill
  \begin{minipage}[b]{0.47\textwidth}
    \includegraphics[width=\textwidth]{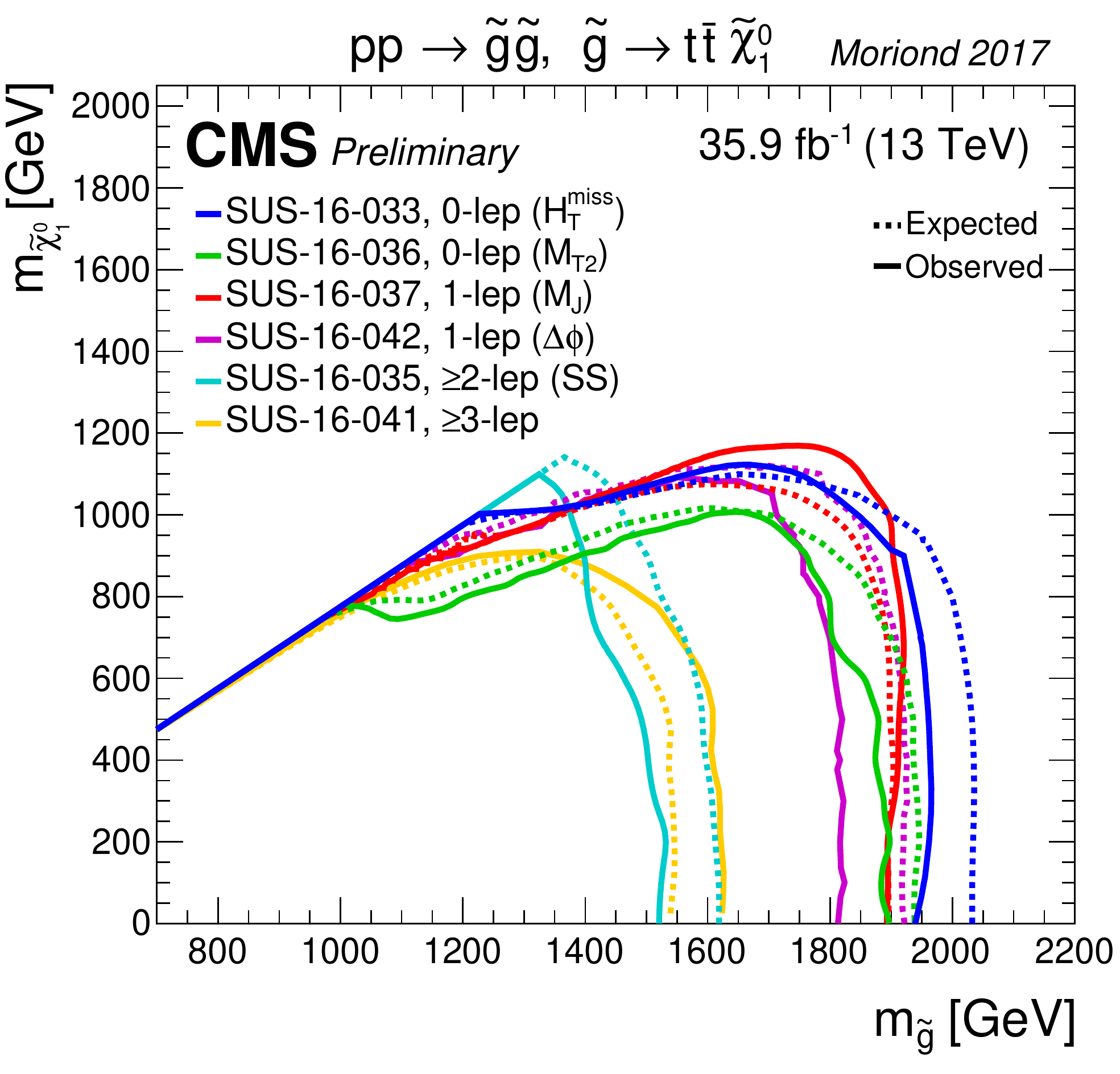}
    \vspace{0.85mm}
    \caption{Summary of limits on gluino pair production with gluino decaying to $t\bar{t}$, computed with 35.9 fb$^{-1}$ of pp collision data taken by CMS at $\sqrt{s}$ = 13 TeV. \label{CMSstrong}}
  \end{minipage}
\end{figure}

After combining a collection of different analyses targeting SUSY strong production, the resulting exclusion contours from both ATLAS (Fig.~\ref{ATLASstrong}) and CMS (Fig.~\ref{CMSstrong}) with about 36 fb$^{-1}$ are obtained and presented in the $m(\tilde{g})-m(\tilde{\chi}^0_1)$ plane. Compared to Run 1, exclusion limits are becoming more and more stringent on the mass of gluinos and of 1$^{st}$ and 2$^{nd}$ generation squarks, and for the first time the sensitivity on masses has reached values beyond 2 TeV.

\section{Direct pair production of 3$^{rd}$ generation squarks}

\begin{figure}[!tbp]
  \centering
  \begin{minipage}[b]{0.45\textwidth}
    \includegraphics[width=\textwidth]{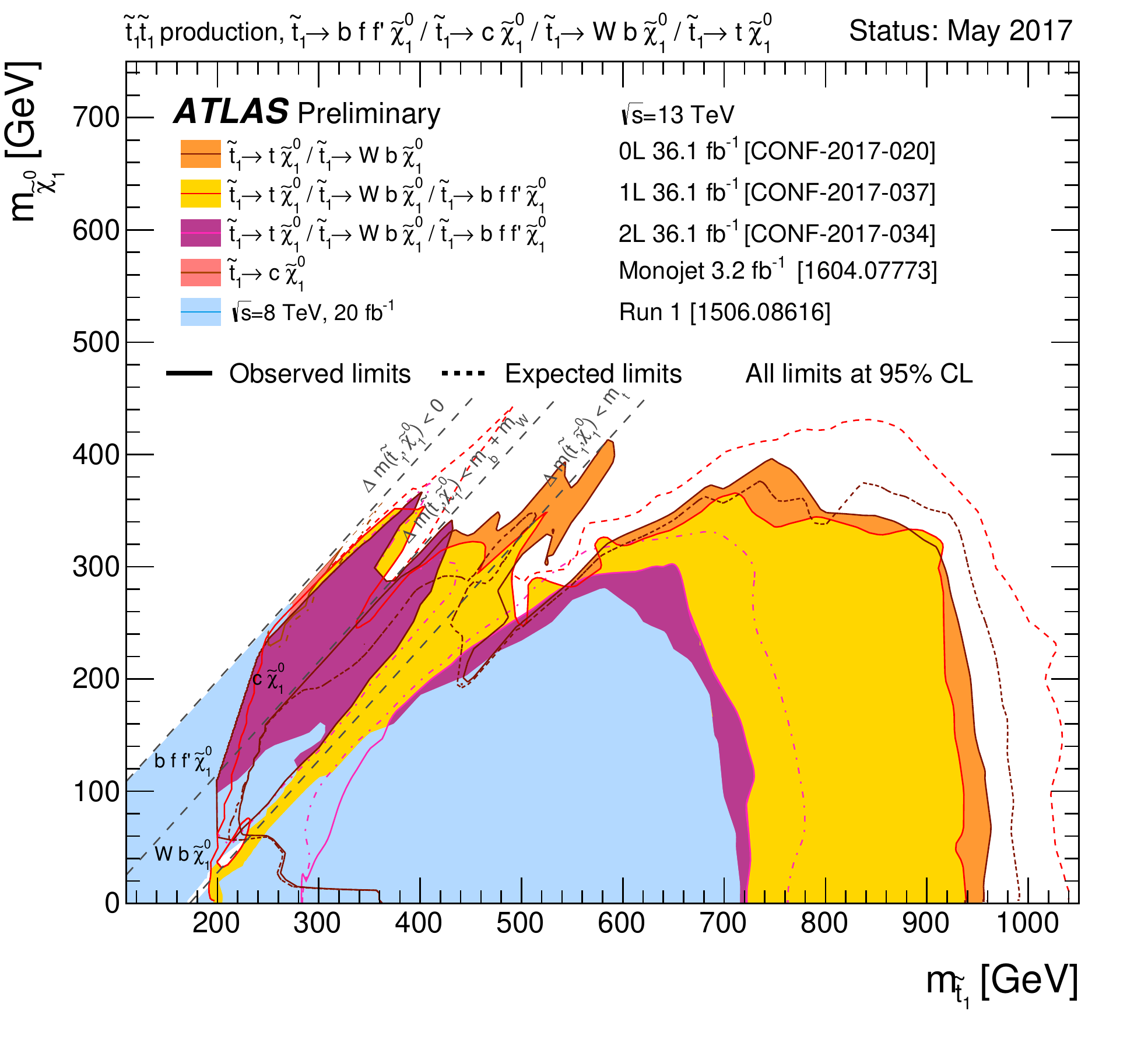}
    \caption{Summary of the dedicated ATLAS searches for top squark pair production based on 3.2 to 36 fb$^{-1}$ of pp collision data taken at $\sqrt{s}$ = 13 TeV. \label{ATLAS3rd}}
  \end{minipage}
  \hfill
  \begin{minipage}[b]{0.45\textwidth}
    \includegraphics[width=\textwidth]{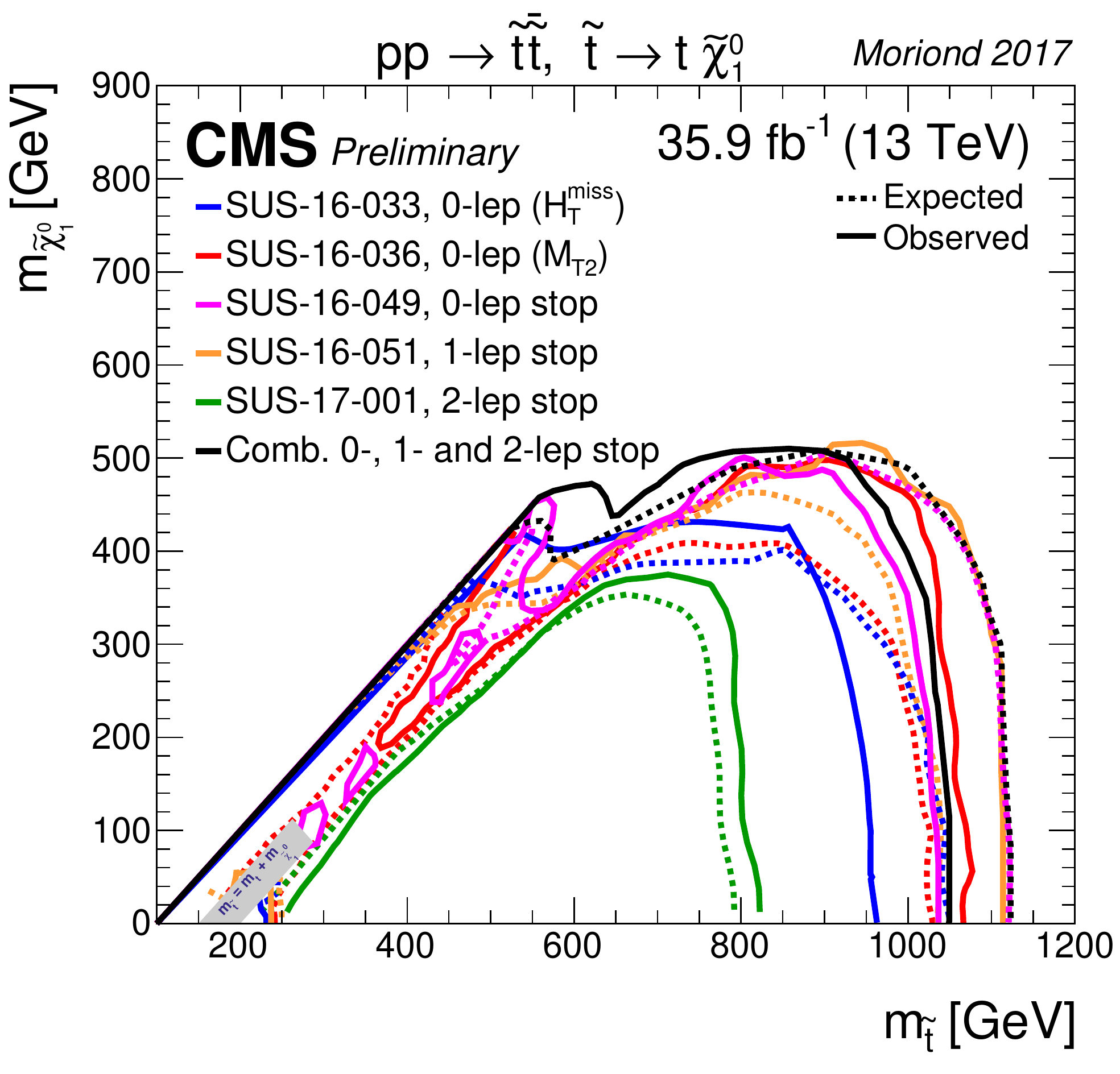}
    \caption{Summary of the dedicated CMS searches for top squark pair production based on 3.2 to 36 fb$^{-1}$ of pp collision data taken at $\sqrt{s}$ = 13 TeV. \label{CMS3rd}}
  \end{minipage}
\end{figure}

The large variety of spectra that the sparticles can have, needs dedicated searches to cover the various possible regions of the $m_{\tilde{\chi}^0_1} - m_{\tilde{t}_1}$ mass plane, with a phenomenology that can be characterized by 2-, 3- or 4- body decays, thus leading to final states with large $E_T^{miss}$, high $b$-jet multiplicity and 0, 1 or 2 isolated leptons.
The summary of limits with $\sim 36$ fb$^{-1}$ for top squark searches is shown for ATLAS and for CMS in Fig.~\ref{ATLAS3rd} and in Fig.~\ref{CMS3rd}, respectively. The exclusion region extends up to more than 1 TeV top squark masses and almost no gaps are left to be further investigated for $m_{\tilde{\chi}^0_1}<$ 400 GeV.

\begin{figure}[pb]
\centerline{\includegraphics[width=6cm]{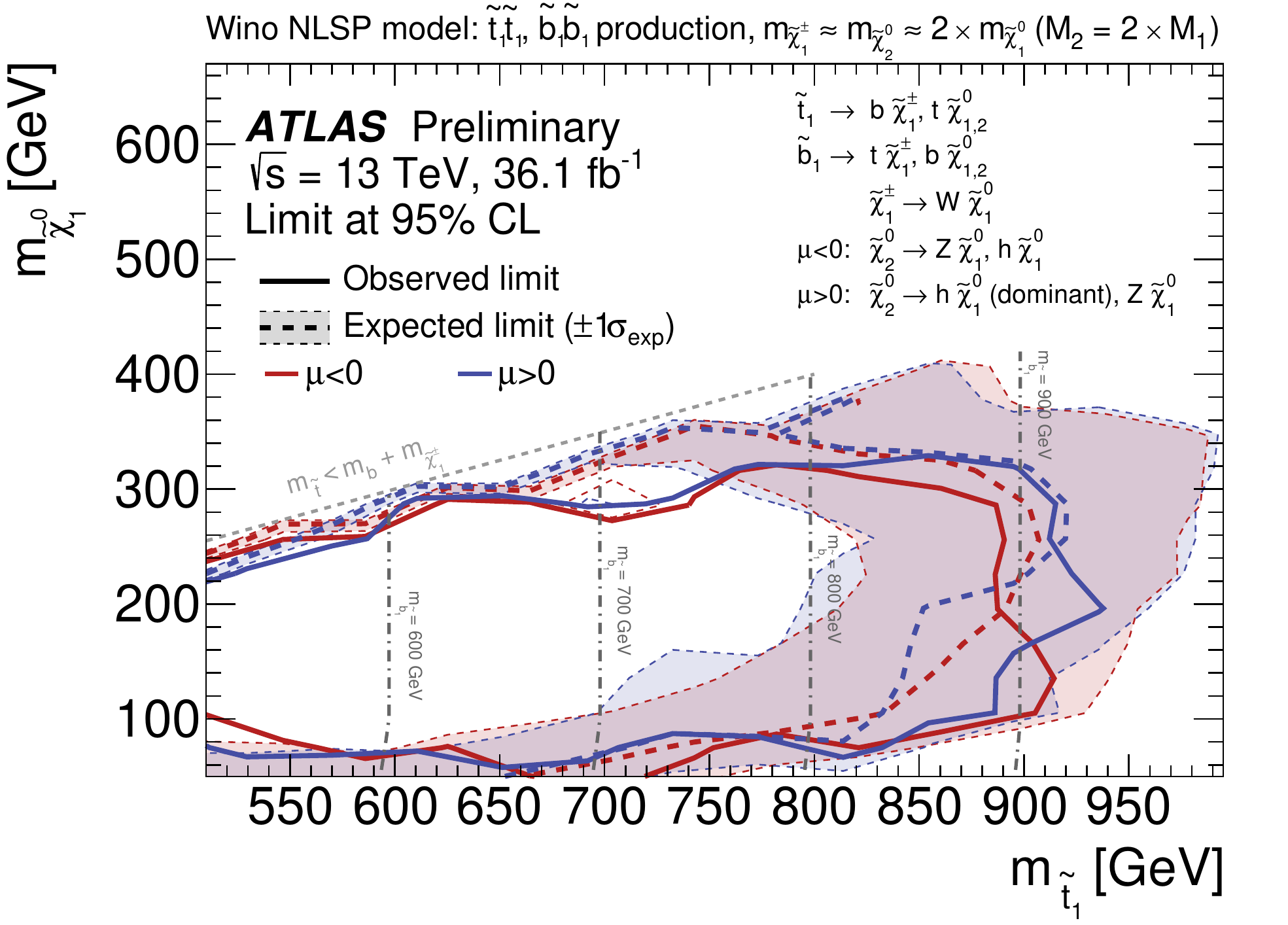}}
\vspace*{8pt}
\caption{Expected and observed excluded regions in the $m_{\tilde{t}_1} - m_{\tilde{\chi}^0_1}$ mass plane for the direct stop/sbottom pair production in the wino NLSP model under the hypothesis of $m_{\tilde{q}_{3L}} < m_{\tilde{t}_R}$, where various decay modes are considered with different branching ratios for each signal point considered by ATLAS.\label{ATLASpMSSM}}
\end{figure}

Beside the simplified $t \tilde{\chi}^0_1$ decay mode, ATLAS searches for phenomenological Minimal Supersymmetric Standard Model (pMSSM) scenarios consider more complex spectra \cite{ATLAS_pMSSM} with other possible $\tilde{t}_1$ decays ($\tilde{t}_1 \rightarrow b \tilde{\chi}^\pm_1$, $\tilde{t}_1 \rightarrow t \tilde{\chi}^0_{1,2}$, $\tilde{b}_1 \rightarrow t \tilde{\chi}^\pm_1$, $\tilde{b}_1 \rightarrow b \tilde{\chi}^0_1$ and $\tilde{b}_1 \rightarrow b \tilde{\chi}^0_2$). As an example, in Fig.~\ref{ATLASpMSSM} expected and observed exclusion limits are shown in the $m_{\tilde{t}_1} - m_{\tilde{\chi}^0_1}$ mass plane for $\tilde{\chi}^\pm_1$ and $\tilde{\chi}^0_2$ having twice the $\tilde{\chi}^0_1$ mass: for $m_{\tilde{\chi}^0_1} \approx$ 200 GeV, the $\tilde{t}_1$ is excluded for masses up to 885 GeV (940 GeV) in scenarios with $\mu < 0$ ($\mu > 0$).

\section{Electroweak production of sparticles}

Supersymmetry can be produced via the electroweak interaction, the direct pair production of charginos and neutralinos being the dominant mode of sparticles production at the LHC in the case in which the SUSY partners of the gluon and of the quarks have a mass heavier than 3-4 TeV.
A huge variety of signatures is tested, mainly exploiting the multi-lepton nature of the final states.

\begin{figure}[!tbp]
  \centering
  \begin{minipage}[b]{0.45\textwidth}
    \includegraphics[width=\textwidth]{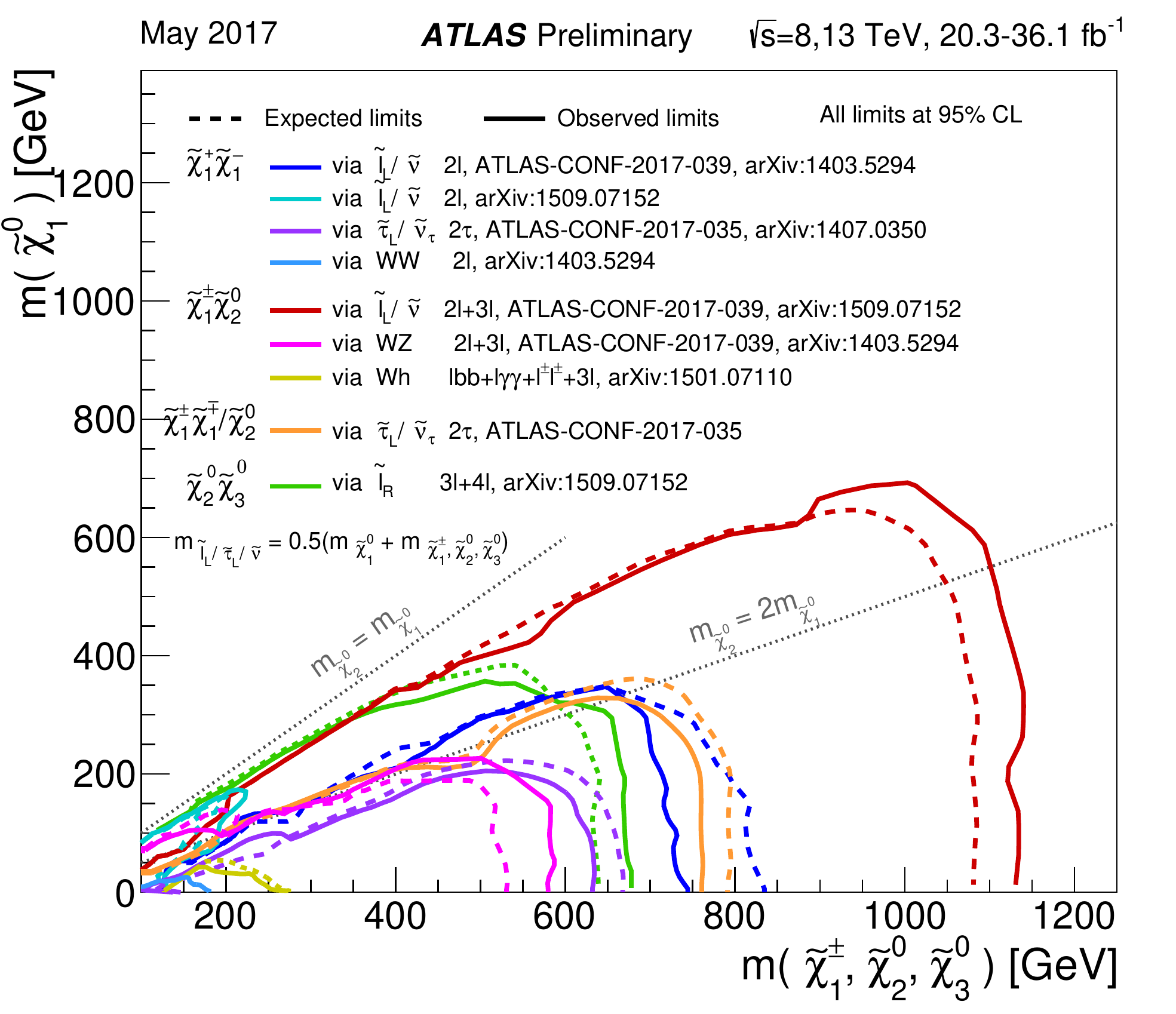}
    \caption{Summary of limits for SUSY electroweak production at ATLAS based on 20.3 to 36.1 fb$^{-1}$ of pp collision data taken at $\sqrt{s}$ = 13 TeV. \label{ATLASEWsum}}
  \end{minipage}
  \hfill
  \begin{minipage}[b]{0.45\textwidth}
    \includegraphics[width=\textwidth]{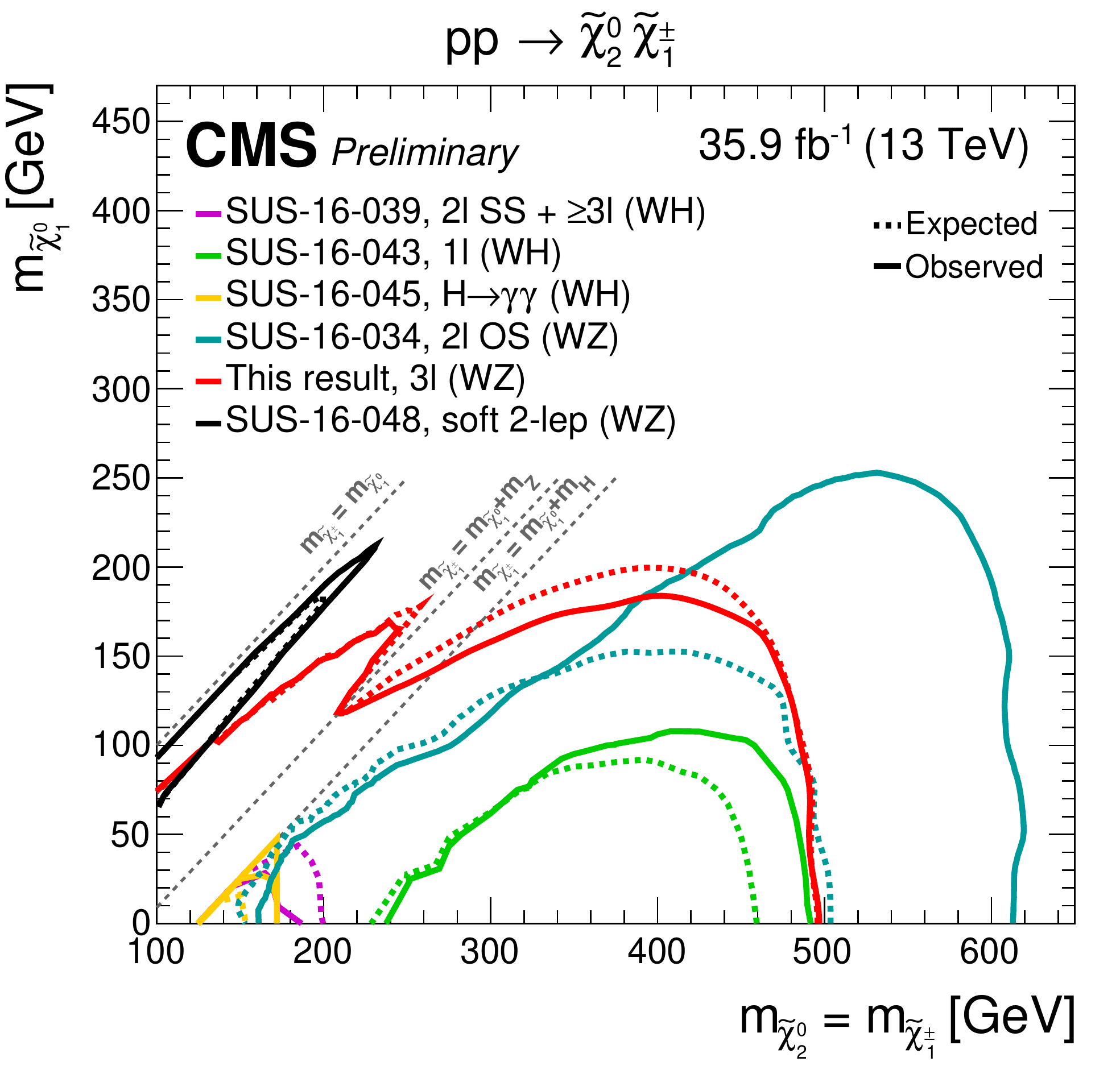}
    \caption{Summary of limits for $pp \rightarrow \tilde{\chi}^0_2 \tilde{\chi}^\pm_1$ production at CMS based on 35.9 fb$^{-1}$ of pp collision data taken at $\sqrt{s}$ = 13 TeV. \label{CMSEWsum}}
  \end{minipage}
\end{figure}

The statistical combination of results for electroweak production searches is reported in Fig.~\ref{ATLASEWsum} for all the channels investigated by ATLAS and in Fig.~\ref{CMSEWsum} for the $\tilde{\chi}^0_2 \tilde{\chi}^\pm_1$ channel.
For decays via $W$, $Z$ and $H$ bosons, the sensitivity is up to $\sim 600$ GeV, while it exceeds 1 TeV for decays via sleptons \cite{ATLASEW}.

\section{R-parity violating scenarios and long-lived particles}
In SUSY there is no theoretical reason why R-parity shouldn't be violated (thus bringing to RPV scenarios). Moreover, whenever mass states are almost degenerate, long-lived particles may exist (for instance in split SUSY and gauge mediated SUSY breaking models). 

\subsection{Searches with one lepton and multi-jet}
\begin{figure}[!tbp]
  \centering
  \begin{minipage}[b]{0.45\textwidth}
    \includegraphics[width=\textwidth]{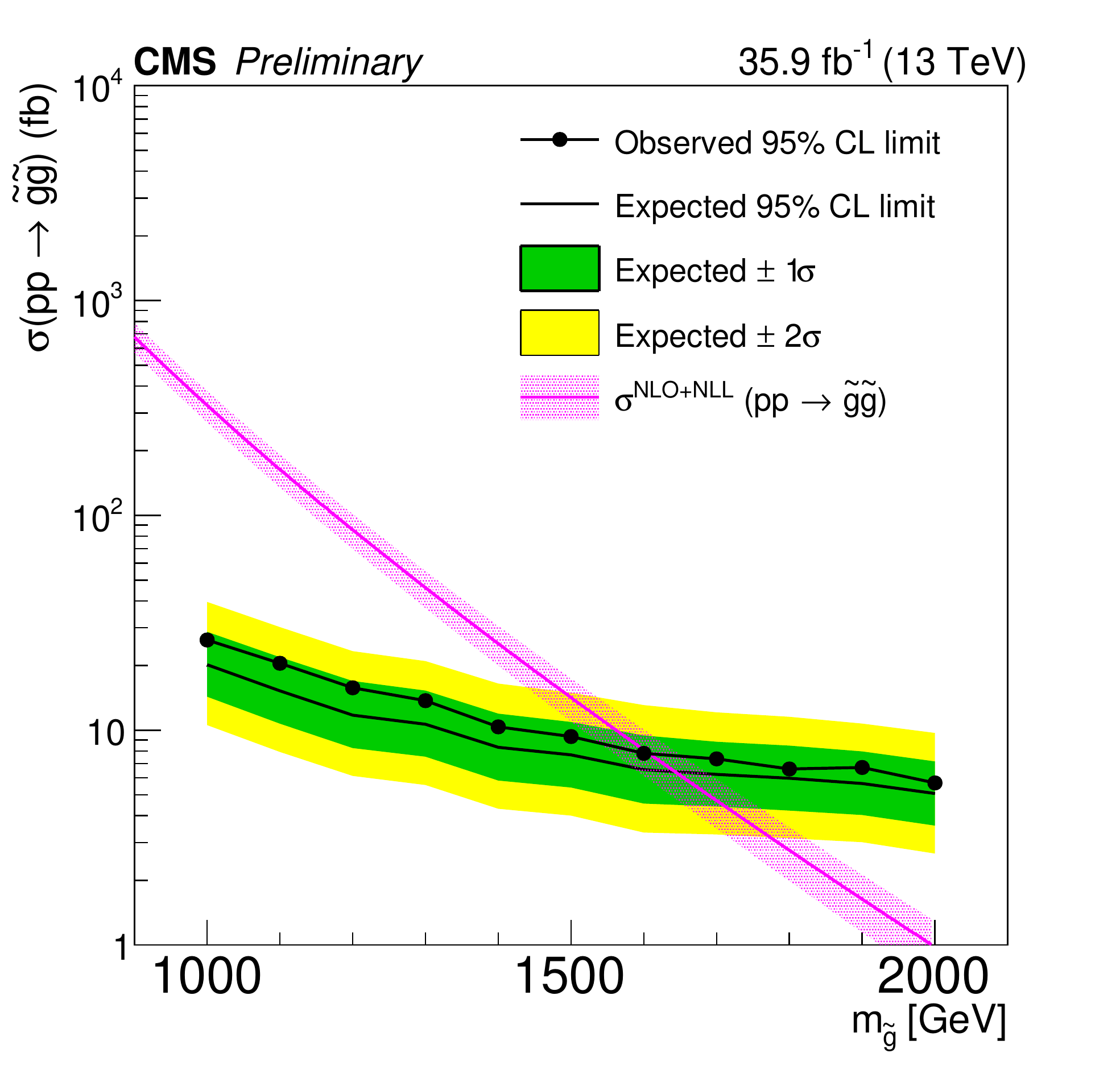}
    \caption{Cross section upper limits at 95\% CL compared to the gluino pair production (magenta), as obtained by CMS with 35.9 fb$^{-1}$ of pp collision data taken at $\sqrt{s}$ = 13 TeV. The theoretical uncertainties are shown as a band around the line. \label{CMSRPV1L}}
  \end{minipage}
  \hfill
  \begin{minipage}[b]{0.45\textwidth}
    \includegraphics[width=\textwidth]{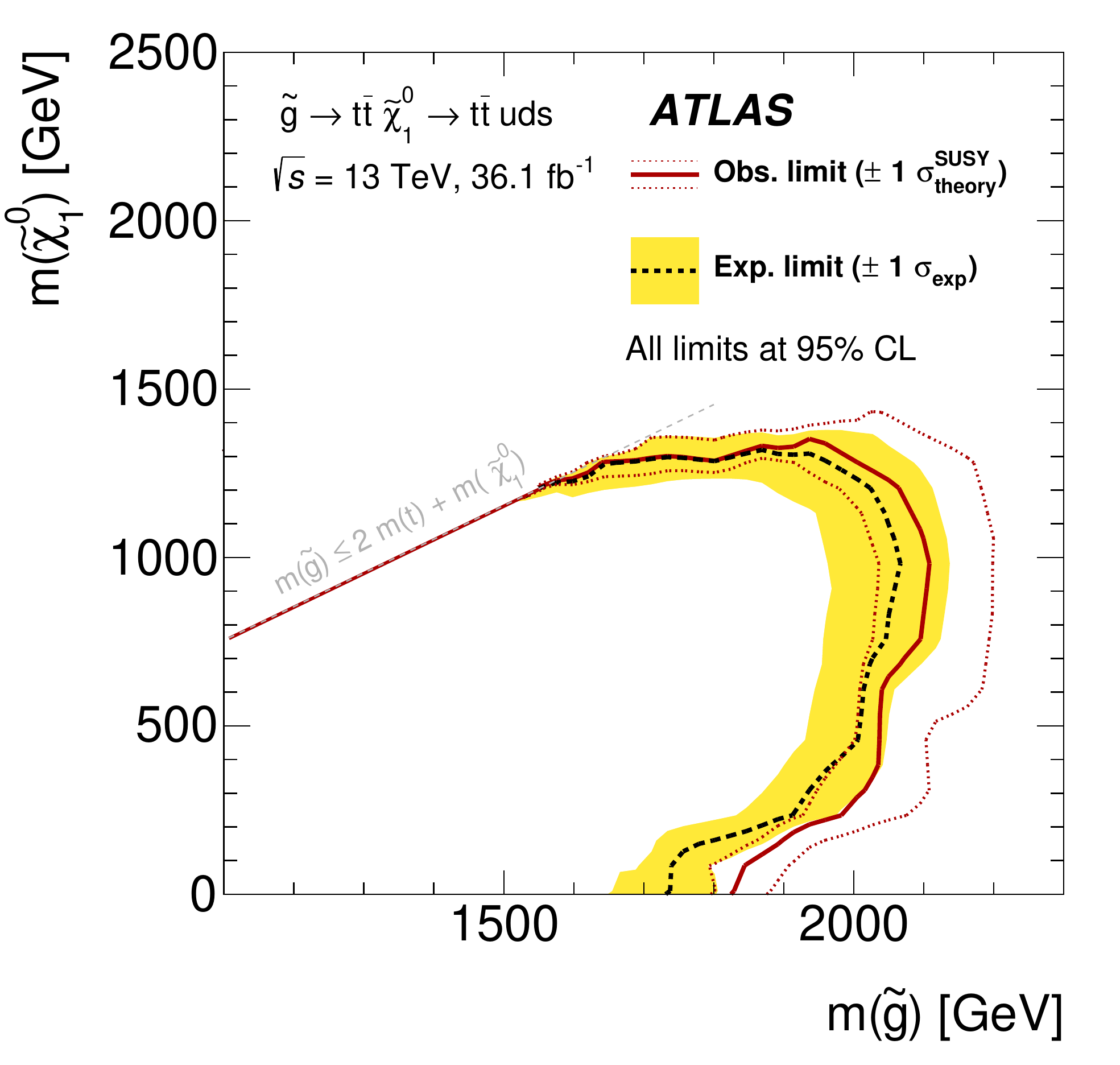}
    \caption{Observed and expected exclusion contours on the $\tilde{g}$ and $\tilde{\chi}^0_1$ or $\tilde{t}$ masses in the context of the RPV SUSY scenarios probed by ATLAS, with simplified mass spectra featuring gluino pair production with exclusive decay modes, based on 36.1 fb$^{-1}$ of pp collision data taken at $\sqrt{s}$ = 13 TeV. \label{ATLASRPV1L}}
  \end{minipage}
\end{figure}

The examples of RPV analyses shown here are carried out by looking for one isolated lepton ($p_T >$ 20 GeV) and at least four jets in the final state, together with limited $E_T^{miss}$.
In Fig.~\ref{CMSRPV1L} results provided by CMS \cite{CMS_RPV1lep} are shown for gluino pair production in a benchmark minimal-flavour-violating model in which gluinos promptly decay to quarks $t$, $b$ and $s$; gluino masses are excluded at 95\% CL for values below 1.61 TeV. 
In the case of ATLAS \cite{ATLAS_RPV1lep}, near fully data-driven background estimations are performed, showing that $t \bar t$ and $W$/$Z$+jets are the dominant source of background. The analysis is based on a model-dependent multi-bin fit which considers both jets and $b$-jets multiplicities.
No significant excess of events is found and lower limits on masses have been set: as shown in Fig.~\ref{ATLASRPV1L}, gluino masses are excluded at 95 \% CL in a range of values between 1.65 and 2.10 TeV, depending on the LSP mass.

\subsection{Long-lived particles}
\begin{figure}[pb]
\centerline{\includegraphics[width=5.4cm]{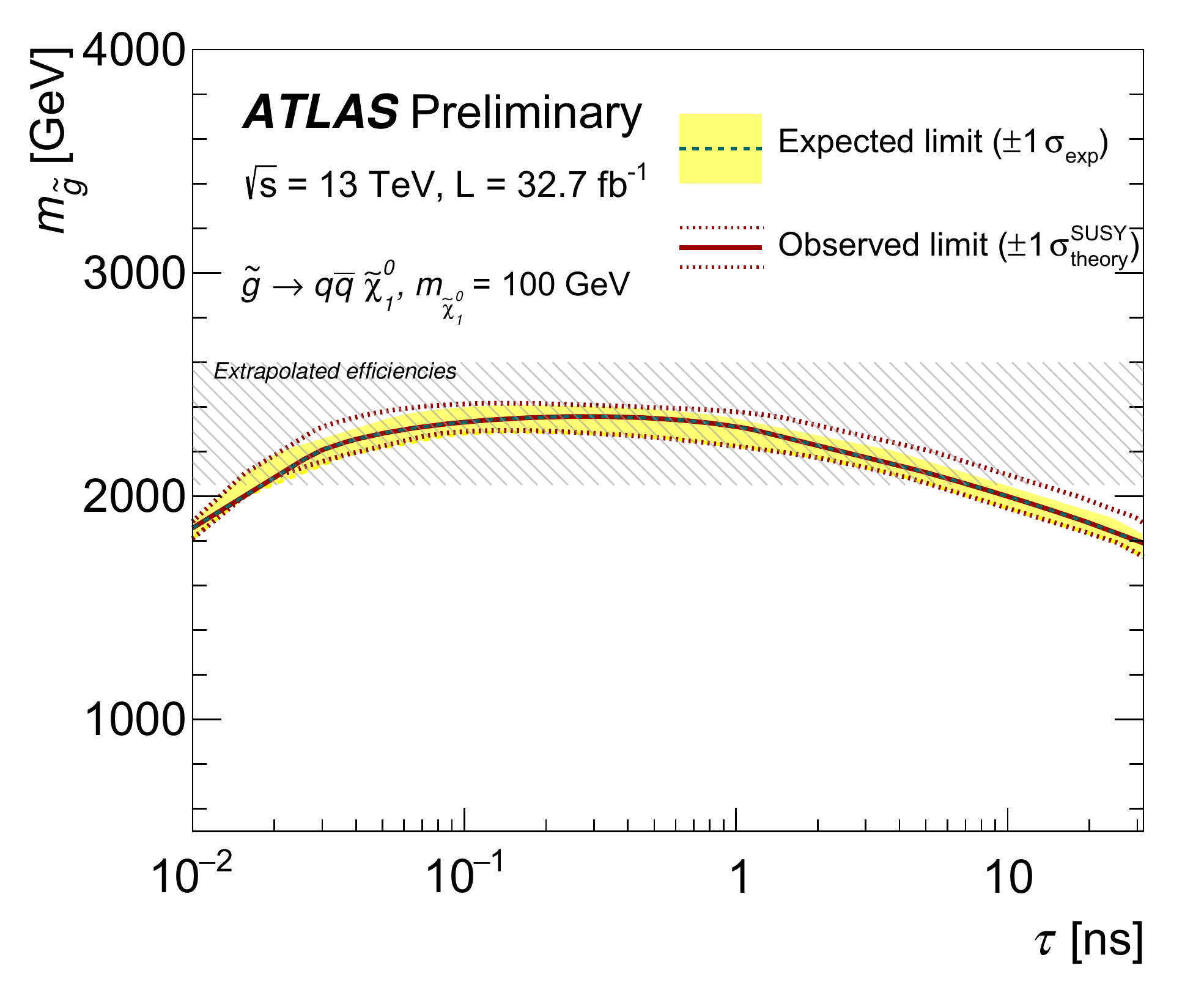}}
\vspace*{8pt}
\caption{Expected and observed upper 95\% CL cross section limits as a function of $\tilde{g}$ lifetime for $m_{\tilde{\chi}^0_1}$ = 100 GeV, based on 32.7 fb$^{-1}$ of pp collision data taken by ATLAS at $\sqrt{s}$ = 13 TeV. \label{ATLASlonglived}}
\end{figure}

Among the various searches for long-lived massive particles, final states with large missing transverse momentum and at least one high-mass displaced vertex with five or more tracks are looked for by ATLAS in the context of split SUSY \cite{ATLAS_longlived}. The considered model foresees very high-mass squarks and a light gluino (at a mass accessible at the LHC) which decays $\tilde{g} \rightarrow q \bar{q} \tilde{\chi}^0_1$ very slowly as it proceeds via a highly virtual squark. The gluino is expected to hadronise and form a bound colour singlet state with SM particles known as an
R-hadrons \cite{Rhadrons}.
In Fig.~\ref{ATLASlonglived} the exclusion limits on the gluino mass are represented as a function of the lifetime in the hypothesis of a 100 GeV neutralino, with masses reaching roughly 2000 to 2300 GeV and lifetimes between 0.02 and 10 ns.

\section{Conclusions}

LHC has performed brilliantly during Run 2, collecting about 36 fb$^{-1}$ of data in years 2015 and 2016, at $\sqrt{s} = 13$ TeV. This has allowed ATLAS and CMS experi\-ments to carry out a wide variety of SUSY searches, for which still no significant deviation from the Standard Model has been observed.
In particular, progresses have been shown regarding searches for strongly produced squarks and gluinos, third generation squarks, direct electroweak production and R-parity violating scenarios. Excluded mass ranges have been extended with respect to LHC Run 1, as they have reached approximately 2 TeV for gluinos, more than 1 TeV for top and bottom squarks and more than 500 GeV for charginos and neutralinos.
Much more data collected at $\sqrt{s} = 13$ TeV in 2017 will enrich the full dataset of Run 2, thus hopefully bringing to new encouraging results on a wide range of SUSY analyses.

\begin{figure}[!tbp]
  \centering
  \begin{minipage}[b]{0.45\textwidth}
    \includegraphics[width=\textwidth]{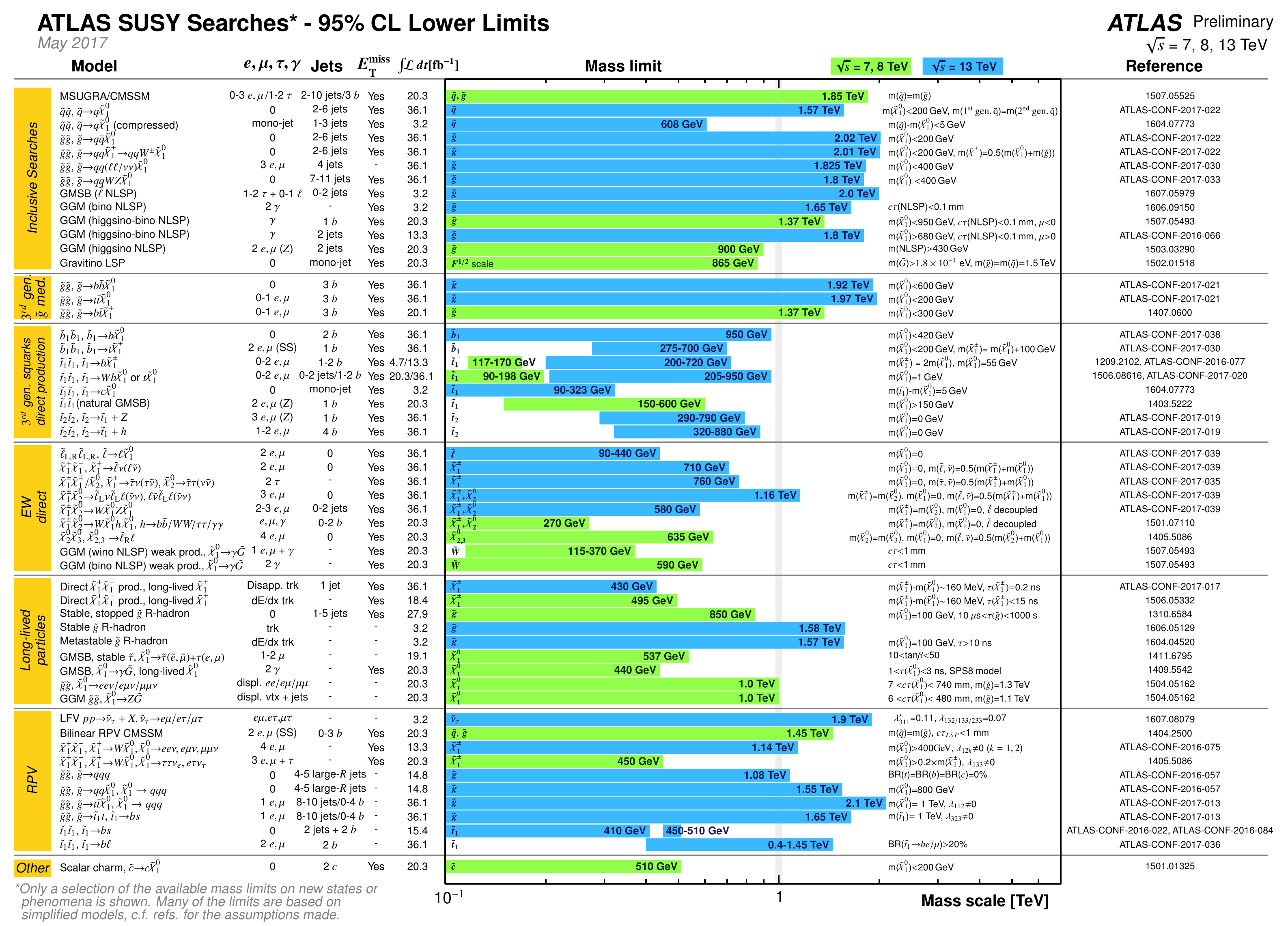}
    \caption{Summary of mass reach of ATLAS searches for Supersymmetry. \label{ATLASsum}}
  \end{minipage}
  \hfill
  \begin{minipage}[b]{0.45\textwidth}
    \includegraphics[width=\textwidth]{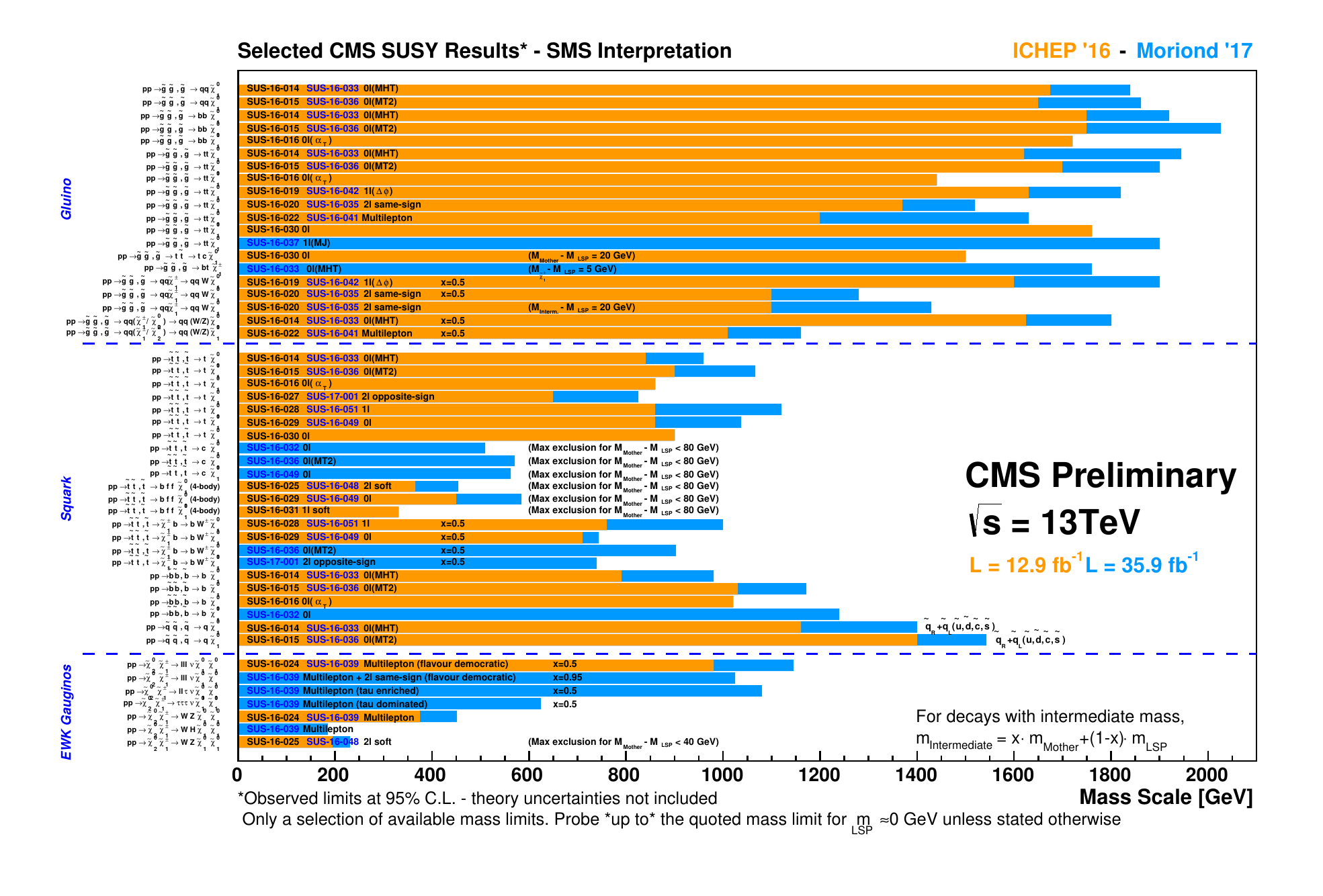}
    \caption{Summary of mass limits obtained by CMS for Supersymmetry. \label{CMSsum}}
  \end{minipage}
\end{figure}
In Fig. \ref{ATLASsum} and in Fig. \ref{CMSsum} a compact representation of the mass limits on sparticles is shown for the ATLAS and for the CMS experiments, respectively, concerning a broad list of different possible scenarios of SUSY production.

\end{document}